\documentclass[12pt]{article}
\usepackage{graphicx}

\newcommand\pubnumber{SNSN-323-63}
\newcommand\pubdate{\today}

\def\institute{Universidad de Oviedo}

\def\Title#1{\begin{center} {\Large #1 } \end{center}}
\def\Author#1{\begin{center}{ \sc #1} \end{center}}
\def\Address#1{\begin{center}{ \it #1} \end{center}}

\newcommand\pubblock{\rightline{\begin{tabular}{l} \pubnumber\\
         \pubdate  \end{tabular}}}
\newenvironment{Abstract}{\begin{quotation}  }{\end{quotation}}
\newenvironment{Presented}{\begin{quotation} \begin{center} 
             PRESENTED AT\end{center}\bigskip 
      \begin{center}\begin{large}}{\end{large}\end{center} \end{quotation}}

\newcommand\ttbar{$t\bar{t}$ }

\def\beq{\begin{equation}}
\def\eeq#1{\label{#1}\end{equation}}
\def\eeqn{\end{equation}}

\def\beqa{\begin{eqnarray}}
\def\eeqa#1{\label{#1}\end{eqnarray}}
\def\eeqan{\end{eqnarray}}

\let\bar=\overbar

\def\Dslash{\not{\hbox{\kern-4pt $D$}}}
\def\dslash{\not{\hbox{\kern-2pt $\del$}}}

\def\msb{{\bar{\ssstyle M \kern -1pt S}}}

\begin{document}
\begin{titlepage}
\pubblock

\vfill
\Title{Measurement of the top quark pair production cross section at $\sqrt{s} = 5.02$ TeV in the dilepton channel with the CMS detector}
\vfill
\Author{ Juan Rodrigo Gonz{\'a}lez Fern{\'a}ndez \\on behalf of the CMS Collaboration}
\Address{\institute}
\vfill
\begin{Abstract}
At hadron colliders, top quarks are dominantly produced in pairs ($t\bar{t}$), a production mechanism that was discovered more than twenty years ago at the Tevatron in Fermilab. Although the \ttbar process has already entered the domain of ``precision'' physics, especially with the advent of the multi-TeV energies at the CERN LHC, there still remains room for dedicated experimental measurements of the top quark in the ``gap'' of energies between Tevatron and LHC. Hence, in the study presented here the total inclusive cross section for pair production at a center-of-mass energy of 5.02 TeV is measured having utilized a low-pileup proton-proton (pp) collision dataset whose recorded luminosity by CMS amounts to 26 pb$^{-1}$. This work, has considered only leptonic W decays, the latter characterized by the presence of an energetic opposite-sign lepton pair plus missing energy. The current measurement can be used for stronger constraints to the poorly known gluon distribution inside the proton at large longitudinal parton momentum fraction, while it paves the way for the very first observation of this elementary particle in nucleus-nucleus collisions. 
\end{Abstract}
\vfill
\begin{Presented}
$9^{th}$ International Workshop on Top Quark Physics\\
Olomouc, Czech Republic,  September 19--23, 2016
\end{Presented}
\vfill
\end{titlepage}
\def\thefootnote{\fnsymbol{footnote}}
\setcounter{footnote}{0}

\section{Introduction and motivation}
The evolution of the inclusive \ttbar production cross section with the center-of-mass energy is of interest for the extraction of the top-quark pole mass and can be used to constrain the gluon distribution function at large longitudinal parton momentum fraction. The fraction of \ttbar events initiated by gluon-gluon collisions grows monotonically with center-of-mass energy and it is around 73\% at 5.02 TeV, to be compared to around 86\% at 13 TeV, making this measurement complementary to the precission measurements at higher center-of-mass energies.

Furthermore, a measurement at 5 TeV will be used as a reference for measurements in nucleon-nucleon collisions at the same center-of-mass energy.

The \ttbar production cross section is measured at 5.02 TeV \cite{top16015} by the CMS detector~\cite{detector} using a dataset that corresponds to $26\pm3$ pb$^{-1}$ of data with low pile-up. The measurement is done using events with an opposite-sign electron-muon pair in the final state, with at least two jets from the hadronization of b-quarks. Due to the low luminosity of the sample and the small branching ratio of this final state the effect of the statistics will completely dominate the uncertainty of the measurement. The cross section is extracted using a event-counting experiment.

\section{Event selection}
Top quarks mostly decay into a b-quark and a $W$ boson, and we will consider final states where both $W$ bosons decay leptonically, having in the final state one opposite-charge electron-muon pair. All particles are reconstructed using the particle flow (PF) algorithm \cite{PF}. The electron (muon) is required to have a $p_{\mathrm{T}}$ of at least $20$ ($18$) GeV and $|\eta| < 2.4$ ($|\eta| < 2.1$) and must be isolated.

The selected dilepton pair must reconstruct an invariant mass greated than $20$~GeV. Selected events must have at least two reconstracted jets. Jets are reconstructed from PF candidates, using an anti-$k_t$ clustering algorithm with a distance parameter of $0.4$. Selected jets must have a $p_{\mathrm{T}}$ of at least $25$ GeV and $|\eta| < 3.0$ at must have a separation from the selected leptons of at least $\Delta R = 0.3$.

Due to the low statistics in our sample we do not require the jets to be identified as containing b-quarks.

\section{Background determination}
Background events are mostly produced from single top (tW) process, Drell-Yan production (DY) or dibosons (WW or WZ). The DY contamination is estimated from data using the $R_{out/in}$ method,  using events with same-flavour leptons to calculate a data-to-MC scale factor, estimated using the number of events in data within a $15$~GeV window around the Z boson mass and extrapolating outside the window using the shape of the distribution from Monte Carlo (MC). The scale factor obtained to normalize the MC simulation is of $0.97\pm0.02$.

Other background sources, such as \ttbar in the lepton + jets final state or W+jets production can contaminate the signal if a jet is incorrectly reconstructed as a lepton, or a lepton is incorrectly identified as being isolated. These events are grouped into a "NonW/Z leptons" category. This instrumental background is estimated using events in a same-sign region in data, with the standard selection but the electron and the muon must have a charge with same sign and the isolation cut for the muon is relaxed to increase the size of the sample.

Other backgounds, such as tW and dibosons, are estimated from MC. The estimated yields can be found in Table \ref{tab-yields}.

\begin{table}[!htb]
\centering
\begin{tabular}{lc}
\hline
         &  Number of \\
Source   & $e^{\pm}\mu^{\mp}$ events      \\
\hline
Drell--Yan              &    1.6 $\pm$   0.4   \\
Non-W/Z leptons         &   1.0 $\pm$   0.9 \\
Single top quark        &   0.89 $\pm$   0.02 \\
WW + WZ                 &    0.41 $\pm$   0.02 \\ \hline
Total background        &   3.9 $\pm$  0.8              \\ \hline
\ttbar\ signal          & 17.0 $\pm$  0.2   \\ \hline
Data                    &    24                   \\
\hline
\end{tabular}
\caption{Number of events obtained after applying the full selection for \ttbar and background estimates and data. The uncertainties correspond to the statistical component.}
\label{tab-yields}
\end{table}

\section{Systematic uncertainties}
The measurement of the cross section is affected by systematic uncertainties originated from several sources but the total effect is negligible in comparison with the statistical uncertanty. The dominant systematic uncertainty corresponds to the muon efficiencies (which includes trigger efficiency). The uncertainty on the luminosity is 12\% and its effect is greater than the combination of all systematic uncertainties. All the uncertainties are summarized in Table \ref{tab-syst}.

\begin{table}[!h]
\centering
\begin{tabular}{lcc}
\hline
Source                        &  $\Delta\sigma_{t\bar{t}}$ (pb)  & $\Delta\sigma_{t\bar{t}} / \sigma_{t\bar{t}}$ (\%) \\
\hline
Electron efficiencies          &  1.1 & 1.4 \\
Muon efficiencies              & 2.4 & 3.0 \\
Jet energy scale              & 1.1 & 1.3 \\
Jet energy resolution         &  0.05 & 0.06 \\
\hline
\ttbar hadronization         &  1.0 & 1.2 \\
Parton shower scale           &  1.0 & 1.2 \\
$\mu_{\rm F}$ and $\mu_{\rm R}$ scales & 0.2 & 0.2 \\
PDF                           &  0.4 & 0.5 \\
\hline
Single top quark              & 1.1 & 1.3 \\
WW + WZ                            & 0.5 & 0.6 \\
Drell--Yan                    & 2.1 & 2.6 \\
Non-W/Z leptons             &  1.9 & 2.3 \\
\hline
Total systematic (no integrated luminosity) & 4.6  &  5.6 \\
Integrated luminosity         & 9.8 & 12.0 \\
Statistical                   &  20 & 24\\
\hline
Total                         & 23  &  28 \\
\hline
\end{tabular}
\caption{Summary of the individual contributions to the uncertainty in the $\sigma_{t\bar{t}}$  measurement. The first and second uncertainty corresponds to the total and relative component, respectively.}
\label{tab-syst}
\end{table}

\section{Results}

The measured cross section is:
\begin{displaymath}
\sigma_{t\bar{t}} = 82 \pm 20~(stat) \pm 5~(syst) \pm 10~(lumi)~pb,
\end{displaymath}
which is consistent within the large uncertainty with the SM prediction of $\sigma_{t\bar{t}} = 68.9^{+1.9}_{-2.3}$ (scale) $ \pm 2.3$ (PDF) $ ^{+1.4}_{-1.0}$ ($\alpha_s$) pb for a top quark mass of 172.5 GeV.



\end{document}